\documentclass[12pt,aps,pra, preprint,english]{revtex4}
\usepackage{graphicx}
\usepackage{dcolumn}
\usepackage{amsmath}
\usepackage{amssymb}
\usepackage{times}
\usepackage{babel}
\usepackage{epstopdf}
\usepackage{color}
\usepackage{float}
\usepackage{epstopdf}

\def\rv{{\bf r}}

\def\qv{{\bf q}}

\def\beq{\begin{equation}}
\def\eeq{\end{equation}}

\begin{document}
\title{Finite temperature analytical results for a harmonically confined gas obeying exclusion statistics in $d$-dimensions}
\author{Zachary MacDonald}
\affiliation{Department of Physics, St. Francis Xavier University, 
Antigonish, NS, Canada B2G 2W5} 
\author{Brandon P. van Zyl}
\affiliation{Department of Physics, St. Francis Xavier University, 
Antigonish, NS, Canada B2G 2W5} 

\date{\today}

\begin{abstract}
Closed form, analytical results for the finite-temperature one-body density matrix, and Wigner function of a $d$-dimensional, harmonically trapped gas of 
particles obeying exclusion statistics are presented.  As an application of our general expressions, we consider the
intermediate particle statistics arising from the Gentile statistics, and compare its
thermodynamic properties to the Haldane fractional
exclusion statistics.  At low temperatures, the thermodynamic quantities derived from both distributions are shown to be in excellent
agreement.  As the temperature is increased, the Gentile distribution continues to provide a good description of the system,
with  deviations only arising well outside of the degenerate regime.  Our results illustrate that the exceedingly simple functional form of the 
Gentile distribution is an excellent alternative to the generally only implicit form of the 
Haldane distribution at low temperatures.
\end{abstract}

\maketitle
\section{Introduction}  

In his now famous 1991 paper, F. D. M.~Haldane~\cite{haldane} proposed a novel generalization of the Pauli exclusion principle, that leads to particle statistics which
continuously interpolate between the Bose and Fermi statistics.  
In the Haldane fractional exclusion statistics (FES), one constructs a generalized exclusion principle through the single particle dimension of the $N$-th particle, in the presence
of $N-1$ other identical particles, {\it viz.,}~\cite{murthy}
\beq
d^{g}_N = d - g(N-1)~,
\eeq
where we note that $g=0$ corresponds to bosons, and $g=1$ to fermions.
The constant $g$ is the so-called FES of a particle, and  by definition is given by
\beq
g = -\frac{\Delta d}{\Delta N}~.
\eeq
In Eq.~(2), $\Delta d$ denotes the change in the dimension of the single particle space, and $\Delta N$ is the change in the number of particles, with
the {\it proviso} that the size and boundary conditions of the system are unchanged.
The FES parameter, $g$, is then a measure of partial Pauli blocking, and can quite generally  take on arbitrary values $g \geq 0$, although we will focus
on $0 \leq g \leq 1$.  
Furthermore, FES is a consequence of state counting arguments made in 
Hilbert space, and so is valid
for arbitrary spatial dimensions.~\cite{khare}

Following Haldane's work, R. Ramanathan~\cite{ramanathan}, Dasin\'eres de Veigy and Ouvry~\cite{ouvry}, Wu~\cite{wu} and Isakov~\cite{isakov} independently examined the thermodynamic properties of an ideal FES gas, and derived the now well-known
result for the average occupancy of a gas of particles obeying ideal FES in the grand canonical ensemble, {\it viz.,}
\beq
n(\epsilon_i) = \frac{1}{f(e^{(\epsilon_i-\mu)/k_BT})+g}~,
\eeq
where $T$ is the temperature, $\epsilon_i$ is the single particle energy,  $k_B$ is the Boltzmann constant,  and $\mu$ is the chemical potential.
The function $f(e^{(\epsilon_i-\mu)/k_BT})$ is determined by ($x=e^{(\epsilon_i-\mu)/k_BT}$)
\beq
f(x)^g(1+f(x))^{1-g} = x~.
\eeq
It is easily seen that for $g=0$, we recover the usual ideal
Bose distribution, while for $g=1$, we obtain the Fermi distribution.   

In general, closed form expressions for $n(\epsilon_i)$ in the Haldane FES, Eq.~(3), are not possible; that is, $n(\epsilon_i)$ is generally only given implicitly through Equation (4).   However,
for the special case of $g= p/q$ where $p$ and $q$ are co-prime, explicit expressions may be obtained.~\cite{aring}  Regrettably, such distributions
quickly become difficult to work with both numerically and analytically.  For example, with $g=1/2$ (semions), and $g = 1/3$, one obtains~\cite{aring}
\beq
n(\epsilon_i) = \frac{2}{\sqrt{1+4 e^{2(\epsilon_i - \mu)/k_BT}}}~,
\eeq
and
\beq
n(\epsilon_i) = \frac{3 }{h + h^{-1}-1}~,
\eeq
\beq
h = [2\sqrt{y(y-1)}+2y - 1]^{1/3}~,~~~~y = 2\left( \frac{3 e^{(\epsilon_i-\mu)/k_BT}}{2}\right)^3+1~,
\eeq
respectively.  It is therefore reasonable to investigate if there is an alternative distribution to Eq.~(3), which is able to 
capture the essential thermodynamic properties of the 
Haldane FES, while still possessing the desirable property of having an {\em explicit}  functional form.

Remarkably, over 70 years ago, the pioneering work of G. Gentile~\cite{gentile} on intermediate particle statistics may have already provided a
possible answer to this question.  Gentile's work was founded on a quantum phase-space approach, in which a single quantum cell may accommodate up to
$1/\alpha_w$ particles with the same energy.  Utilizing  the method of most probable distribution, Gentile obtained
what we shall refer to as the Gentile exclusion statistics (GES) distribution, {\it viz.,} 
\beq
n^{(\alpha_w)}(\epsilon_i) = \frac{1}{e^{(\epsilon_i-\mu)/k_BT} - 1} - \frac{ w}{e^{w(\epsilon_i-\mu)/k_BT} - 1}~,
\eeq
where $w\equiv \frac{1+\alpha_w}{\alpha_w}$.  It is clear that for $\alpha_w=0$ and $\alpha_w=1$, Eq.~(8) exactly reproduces the standard Bose and Fermi distributions, respectively.  
More importantly, 
Eq.~(8) also provides a simple, form invariant,  expression for the average occupation number, 
$n^{(\alpha_w)}(\epsilon_i)$, $\forall$ $0\leq \alpha_w \leq 1$.  
We wish to point out that Eq.~(8) appears to have been recently ``rediscovered'' by Q. A. Wang {\it et. al}~\cite{wang}, who based their analysis on the grand 
partition function, also supposing 
$n_{\rm max}=1/\alpha_w$ for the maximum occupation number.
In fact, the distribution derived in Ref.~\cite{wang} is identical to what is obtained in Gentile statistics, Eq.~(8), although Wang {\it et al.} do not appear to be 
aware of this fact.~\cite{gentile,nanda,sarkar}

One of the objectives of this paper is to investigate the viability of the GES explicit 
distribution, {\it viz.,} Eq.~ (8), as an
alternative to the {\em implicit} FES distribution given by Equations (3) and (4).  We note that this is a meaningful comparison, since both GES and FES are rooted in 
the generalization of the Pauli principle, with a well defined method for the counting of states.~\cite{note1}
In order to facilitate this goal, we will focus our attention to a $d$-dimensional gas of ideal particles obeying arbitrary statistics at finite-temperature,
confined to a harmonic oscillator trap, $V(R)=m\omega^2 R^2/2$, with $R=\sqrt{x_1^2+x_2^2+\cdot\cdot\cdot+x_d^2}$ being the $d$-dimensional hyper-radius.  The motivation for studying this system lies in its possible
connection to current experiments on harmonically trapped, ultra-cold quantum gases, along with the models relatively simple analytical properties.

To this end, the rest of our paper is organized as follows.  In Sec.~II, we will present finite temperature,
closed form analytical expressions for the $d$-dimensional one-body density matrix (ODM) and Wigner function
obeying general exclusion statistics.  
Then, in Sec.~III, we make use of the Wigner function to construct a variety of thermodynamic properties
without restriction to any specific  statistics.  In Sec.~IV, we narrow our focus to  FES and GES, so that we may make a
detailed comparison of these two distributions at finite, and zero-temperature.  In Sec.~V we present
our concluding remarks.

\section{Finite temperature one-body density matrix and Wigner function}
In this section, we will provide closed form expressions for the finite-temperature 
ODM and Wigner function obeying
exclusion statistics, in arbitrary dimensions for a harmonically trapped gas.  In what follows, we will denote the general exclusion statistics
parameter by $\alpha$, such that $0< \alpha\leq 1$ defines the fermionic sector.  This is in fact
a generic feature of any exclusion statistics distribution, which is required to continuously interpolate between Bose and Fermi statistics (see also Eq.~(15) below).
The finite-temperature
Wigner function is subsequently used to evaluate a variety of thermodynamic properties,
such as the spatial density, momentum density, kinetic energy density, and form factor.  
\subsection{One-body density matrix}
The $d$-dimensional, finite-temperature ODM, $\rho^{(d)}(r,s;T)$, for a system obeying
{\em arbitrary} statistics is obtained by taking the two-sided inverse Laplace transform (ILT) of the finite-temperature Bloch-density matrix,~\cite{brack_bhaduri} 
\begin{equation}
\rho^{(d)}(r,s;T) = {\cal B}_{\mu}^{-1}\left[  C^{(d)}_T(r,s;\beta)\right]~,
\end{equation}
where
\beq
C^{(d)}_T({r},{ s};\beta) = C_0^{(d)}( r , s; \beta) \times \frac{H^{(\alpha)}(\beta;T)}{\beta}~,
\eeq
and $\beta$ is the generally complex variable conjugate to $\mu$.  We have also introduced the center-of-mass and relative coordinates 
\beq
{\bf r} = \frac{{\bf r}_1+{\bf r}_2}{2},~~~~~{\rm and}~~~{\bf s} = {\bf r}_1 - {\bf r}_2~,
\eeq 
respectively.  Hereby, we shall use units such that $k_B=\hbar=m=\omega=1$.  
The factor, $H^{(\alpha)}(\beta;T)/\beta$, in Eq.~(10) is a thermal weighting factor, taking into account the statistics, and $C^{(d)}_0({r},{ s};\beta)$ is the normal ($T=0$) Bloch-density matrix.~\cite{brack_bhaduri}

The ILT in Eq.~(9) will be a convolution between the $\beta$-dependence of $C^{(d)}_0({r},{ s};\beta)$, and $H^{(\alpha)}(\beta;T)/\beta$.  
Therefore, it would appear that the thermal factor must be known
explicitly in order to obtain the ODM of the system.  
However, we now point out the  following defining property of the two-sided ILT for {\em any} given thermal factor $H^{(\alpha)}(\beta;T)/\beta$, which we  write as
\beq
{\cal B}^{-1}_{\mu} \left[\frac{H^{(\alpha)}(\beta;T)}{\beta}\right] = n^{(\alpha)}(0)~.
\eeq
In Eq.~(12), $n^{(\alpha)}(\epsilon)$ is the distribution associated with the statistics of the particles.  
For example, let us consider the Bose and Fermi distributions
\beq
n_B(\epsilon) = \frac{1}{e^{(\epsilon -\mu)/T} -1} ~,
\eeq
and
\beq
n_F(\epsilon) = \frac{1}{e^{(\epsilon -\mu)/T} +1} ~,
\eeq
respectively.  Generalizing the above cases, one may write~\cite{polychrono}
\beq
n^{(\alpha)}(\epsilon) = \frac{1}{e^{(\epsilon-\mu)/T} + \alpha}~,
\eeq
so that Bose ($\alpha=-1$), Boltzmann ($\alpha=0$), and Fermi ($\alpha=1$) statisics are all represented by a universal distribution.
Focusing on the Bose and Fermi distributions,  explicit expressions for the
thermal factors $H^{(\alpha)}(\beta;T)/\beta$ are known; namely, for bosons~\cite{vanzyl_bhaduri}
\beq
\frac{H_B(\beta;T)}{\beta} = - \frac{\pi T}{\tan(\pi \beta T)}~,
\eeq
whereas for fermions,
\beq
\frac{H_F(\beta;T)}{\beta} = \frac{\pi T}{\sin(\pi \beta T)}~.
\eeq
It can be readily shown by direct calculation that~\cite{vanzyl_bhaduri}
\beq
{\cal B}^{-1}_{\mu} \left[\frac{H_B(\beta;T)}{\beta}\right] = \frac{1}{\exp(-\mu/T) - 1} = n_B(0)~,
\eeq
and
\beq
{\cal B}^{-1}_{\mu} \left[\frac{H_F(\beta;T)}{\beta}\right] = \frac{1}{\exp(-\mu/T) +1} = n_F(0)~.
\eeq
Therefore, once the $\beta$-dependence of $C^{(d)}_0({r},{ s};\beta)$ is known, an application of the convolution theorem for Laplace transforms will, at least in principle, be able to provide us
with the finite-temperature ODM {\it via} Equation (9).  
Note that the convolution integral may still be very difficult to evaluate analytically if the $\beta$-dependence coming from the normal
Bloch-density matrix is complicated.  Indeed, the analytical evaluation of the ILT may only be feasible for specific dimensions.

However, we now make a critical observation.  If $C^{(d)}_0({r},{ s};\beta)$ can be written such that the $\beta$-dependence is exponential, the shifting property of the 
Laplace transform~\cite{grad} may be used
to find a universal expression for the finite-temperature ODM, which is unchanged by the dimension or statistics under consideration.   To wit, we note that
by the shift property, we have
\beq
{\cal B}^{-1}_{\mu} \left[e^{-\gamma \beta} \frac{H^{(\alpha)}(\beta;T)}{\beta}\right] = n^{(\alpha)}(\gamma)~,
\eeq
where $\gamma$ is real and positive.  In other words, if the Bloch density matrix is purely exponential in its $\beta$-dependence, the ODM may be found for {\em any} statistics without
requiring an explicit expression for the thermal factor, $H^{(\alpha)}(\beta;T)/\beta$.
It is  then highly desirable to try to express the $\beta$-dependence of $C^{(d)}_0({r},{ s};\beta)$ as a pure exponential.  This goal is actually achievable for the case of  $d$-dimensional
harmonic confinement, where we
obtain the following expression for the normal Bloch-density matrix, {\it viz.,}~\cite{shea_vanzyl}
\begin{equation} \label{eq3}
C_0^{(d)}( r ,  s; \beta) = \frac{g_s}{\pi^{d/2}}\sum_{n=0}^{\infty}\sum_{k=0}^{\infty} (-1)^n L_n^{d/2-1}(2r^2)L_k^{d/2-1}(s^2/2)e^{-(r^2+s^2/4)}e^{-(\epsilon_n+k)\beta}~,
\end{equation}
where $\epsilon_n = (n+d/2)$ is the $d$-dimensional spectrum of an isotropic harmonic oscillator potential, and plays the role of $\gamma$ in Equation (20).  The quantity, $g_s$ in 
Eq.~(21) denotes the spin degeneracy, and $L_n^a(x)$ are the associated Laguerre polynomials.~\cite{grad}

It then immediately follows that the ILT in Eq.~(9) may be performed {\em without requiring} an explicit expression for the thermal factor, by using the general result
\beq
{\cal B}^{-1}_{\mu} \left[ e^{-(\epsilon_n+k)\beta} \frac{H^{(\alpha)}(\beta;T)}{\beta}\right] = n^{(\alpha)}(\epsilon_n+k)~.
\eeq

In order to clarify, and illustrate the above analysis, let us again consider the Bose and Fermi statistics, from which
Eqs.~(18) and (19) provide us with the appropriate ILTs for the thermal factors, $H^{(\alpha)}(\beta;T)/\beta$.   We may evaluate the ILT piece in Eq.~(9) for bosons by brute force using the
convolution theorem for ILTs, {\it viz.,}
\begin{eqnarray}
{\cal B}^{-1}_\mu \left[e^{-(\epsilon_n+k)\beta} \frac{- \pi T}{\tan(\pi \beta T)}\right] &=& \int_{-\infty}^{\infty} d\tau ~\delta(\tau - (\epsilon_n+k))\frac{1}{\exp[(\tau - \mu)/T]-1}\nonumber \\
&=& \frac{1}{e^{(\epsilon_n+k - \mu)/T}-1}\nonumber \\
&\equiv& n_B(\epsilon_n+k)~,
\end{eqnarray}
and similarly for fermions
\begin{eqnarray}
{\cal B}^{-1}_\mu \left[e^{-(\epsilon_n+k)\beta} \frac{\pi T}{\sin(\pi \beta T)}\right] &=& \int_{-\infty}^{\infty} d\tau ~\delta(\tau - (\epsilon_n+k))\frac{1}{\exp[(\tau - \mu)/T]+1}\nonumber \\
&=& \frac{1}{e^{(\epsilon_n+k - \mu)/T}+1}\nonumber \\
&\equiv& n_F(\epsilon_n+k)~.
\end{eqnarray}
Notice that a direct application of Eq.~(20) immediately leads to the same result, without explicit knowledge of $H^{(\alpha)}(\beta;T)/\beta$.

We may then write down the general expression for the finite-temperature ODM of a harmonically trapped gas, appropriate for general exclusion statistics, as
\begin{eqnarray}
 \rho^{(d)}( r,  s;T)& = &\frac{g_s}{\pi^{d/2}}\sum_{n=0}^{\infty}\sum_{k=0}^{\infty} (-1)^n L_n^{d/2-1}(2r^2)L_k^{d/2-1}(s^2/2)e^{-(r^2+s^2/4)} n^{(\alpha)}(\epsilon_n+k)~.
\end {eqnarray}
Thus, for bosons, Eq.~(25) would read exactly as above, but with $n^{(\alpha)}(\epsilon_n+k) \rightarrow n_B(\epsilon_n+k)$.  Similarly, for fermions, Eq.~(25) still holds, but with $n^{(\alpha)}(\epsilon_n+k) \rightarrow n_F(\epsilon_n+k)$.
We must emphasize that this universal expression for the finite-temperature ODM is only possible owing to the special decomposition of $C^{(d)}_0({r},{ s};\beta)$, such that the $\beta$ dependence is
strictly exponential.  This is by no means a trivial result, and would be difficult, if not impossible to establish by starting with the single-particle harmonic oscillator 
eigenstates in $d$-dimensions.  In fact, for any other form of
$C^{(d)}_0({r},{ s};\beta)$, the temperature dependence in Eq.~(25) changes with dimensionality, and the universal representation of the ODM is lost,
as illustrated in Reference~\cite{vanzyl}.  In addition, note that in Eq.~(25), both the center-of-mass, $r$, and relative coordinate, $s$, are treated on equal footing,  resulting in a clean separation of the variables.  This form
for the ODM is useful for analytical calculations where separate integrations over $r$ and $s$ may need to be performed.
\subsection{Wigner function}
The $d$-dimensional Wigner function, $W^{(d)}(r,p;T)$,  may now be obtained {\it via} a Fourier transform of the ODM, Eq.~(25), with respect to the relative coordinate.  
The Wigner function is a useful tool for the phase-space formulation of quantum mechanics,~\cite{wigner} and as we shall see below, also simplifies analytical calculations
for various thermodynamic properties of the system.
Specifically, by definition, 
\begin{equation}\label{defwig}
W^{(d)}(r,  p;T) = \int d^d s ~\rho^{(d)}( r , s; T) e^{-i \mathbf p \cdot  \mathbf s}~,
\end{equation}
where $d^d x = 2 \pi^{d/2} x^{d-1}/\Gamma[d/2]$. 
Given that the ODM only depends on the magnitude of the coordinates, all angular integrals may be immediately performed, thereby allowing us to write Eq.~(26) as~\cite{vanzyl_wigner}
\begin{equation} \label{wigJ}
W^{(d)}(r, p;T) = (2\pi)^{d/2} \int_0^{\infty} ds~\rho^{(d)}(r,s;T)\bigg(\frac{1}{ps}\bigg)^{d/2-1} J_{\frac{d}{2}-1}(sp)s^{d-1}~,
\end{equation}
where $J_{n}(x)$ is a Bessel function of the first kind.~\cite{grad}  The integral in Eq.~(27) has already been addressed in an earlier work,~\cite{vanzyl_wigner} 
and following the same analysis, we readily obtain the desired result
\begin{eqnarray} \label{fbwig}
W^{(d)}( r,  p;T) &=& 2^dg_s \sum_{n=0}^{\infty}\sum_{k=0}^{\infty}(-1)^{n+k}L_n^{d/2-1}(2r^2)L_k^{d/2-1}(2p^2)e^{-(r^2+p^2)}n^{(\alpha)}(\epsilon_n+k)~.
\end{eqnarray}
Similar to  Eq.~(25), there is once again a clean separation 
of the variables in Eq.~(28), which in this case are the spatial and momentum variables.  The utility of this form for the Wigner function will be illustrated below.

The finite temperature expressions given by Eqs.~(25) and (28) are valid for any dimensionality, any flavour of exclusion statistics, and represent the main analytical 
results of this paper.
\section{Finite temperature results}
Here, we make use of the Wigner function developed above  to construct several thermodynamic quantities of interest.  The results presented here
serve to generalize the Bose and Fermi expressions presented elsewhere in the literature.~\cite{vanzyl_wigner}
\subsection{Spatial density}
The spatial density is obtained from the Wigner function {\it via}
\begin{eqnarray}
\rho^{(d)}(r;T) &=& \frac{1}{(2\pi)^d} \int d^d p ~W^{(d)}(r,p;T)~.
\end{eqnarray}
Using the integral~\cite{grad}
\begin{eqnarray}
I = \int_0^\infty dx~L_k^{d/2-1}(2 x^2) e^{-x^2} x^{d-1} =\frac{1}{2}(-1)^k \frac{\Gamma(k+d/2)}{\Gamma(k+1)}~,
\end{eqnarray}
Eq.~(29) evaluates to
\begin{equation}
\rho^{(d)}(r;T) = 
\frac{g_s}{\pi^{d/2}} \sum_{n=0}^{\infty}\sum_{k=0}^{\infty} (-1)^n 
\left(
\begin{array}{cc}
k+d/2-1 \\
k
\end{array}
\right)
L_n^{d/2-1}(2 r^2)e^{-r^2} n^{(\alpha)}(\epsilon_n +k)~.
\end{equation}
Observe that the convenient separation of the $r$ and $p$ coordinates in Eq.~(28) has allowed for an easy calculation of the spatial density.  Of course, Eq.~(31) may also be
obtained by setting $s=0$ in Equation (25).
\subsection{Momentum density}
The finite temperature momentum density, $\Pi^{(d)}(p;T)$, is obtained by integrating over the coordinate variable, {\it viz.,}
\begin{eqnarray}
\Pi^{(d)}(p;T) &=& \frac{1}{(2\pi)^d} \int d^d r ~W^{(d)}(r,p;T)~.
\end{eqnarray}
Once again, making use of Eq.~(30), we obtain
\begin{equation}
\Pi^{(d)}(p;T) = 
\frac{g_s}{\pi^{d/2}} \sum_{n=0}^{\infty}\sum_{k=0}^{\infty} (-1)^k 
\left(
\begin{array}{cc}
n+d/2-1 \\
n
\end{array}
\right)
L_k^{d/2-1}(2 p^2)e^{-p^2} n^{(\alpha)}(\epsilon_n+k)~.
\end{equation}

\subsection{Kinetic energy density}
The finite temperature kinetic energy density, $\tau^{(d)}(r;T)$,  is calculated according to

\begin{equation}
\tau^{(d)}(r;T) = \frac{1}{(2\pi)^d}\int_0^\infty d^dp \text{ }W^{(d)}(r,p;T) \frac{p^2}{2}~.
\end{equation}
Inserting the finite-temperature Wigner function, Eq.~(28), into Eq.~(34), and performing the integration, leads to 

\begin{equation} \label{keden_T}
\tau^{(d)}(r;T) = \frac{g_s }{4 \pi^{d/2} \Gamma(d/2)} \sum_{n=0}^{\infty}\sum_{k=0}^{\infty}(-1)^{n} \frac{(d+4k)\Gamma(d/2+k)}{k!}  L_n^{d/2-1}(2r^2)e^{-r^2} n^{(\alpha)}(\epsilon_n+k)~.
\end{equation}
From here, the kinetic energy can be obtained by integrating Eq.~\eqref{keden_T} over all space,
\begin{equation}
E_{\text{kin}}(T)  = \int_0^{\infty} d^dr \text{ } \tau^{(d)}(r;T).
\end{equation}
Using tabulated integrals,~\cite{grad} it is straightforward to show that 

\begin{equation}
E_{\text{kin}}(T) =  \frac{g_s }{4 \Gamma(d/2)^2}  \sum_{n=0}^{\infty}\sum_{k=0}^{\infty}\frac{(d+4k)\Gamma(d/2+n) \Gamma(d/2+k)}{n!k!} n^{(\alpha)}(\epsilon_n+k)~.
\end{equation}

\subsection{Form factor}
The form factor is simply the Fourier transform of the spatial density, and is given by, 
\beq
f^{(d)}(q;T) = \int d^dr~\rho^{(d)}(r;T) e^{-i\qv\cdot\rv}~.
\eeq
Following an identical analysis as for the evaluation of $W^{(d)}(r,p;T)$ yields~\cite{vanzyl_wigner}
\begin{eqnarray}
f^{(d)}(q;T) &= &
g_s\sum_{n=0}^{\infty}\sum_{l=0}^{\infty} 
\left(
\begin{array}{cc}
l+d/2-1 \\
l
\end{array}
\right)
L_n^{d/2-1}(q^2/2)e^{-q^2/4}~n(\epsilon_n + l)\nonumber \\
&=& g_s \sum_{m=0}^\infty L_m^{d-1}(q^2/2) e^{-q^2/4}~n^{(\alpha)}(\epsilon_m)~.
\end{eqnarray}

\subsection{Zero temperature}

In the case of the {\em fermionic branch}, the $T\to 0$ limit is easily obtained by noting that
\beq
n^{(\alpha)}(\epsilon_n+k) \to \frac{1}{\alpha} \Theta (\epsilon_f^{(\alpha)} - (\epsilon_n+k)),
\end{equation}
where the Fermi energy is now given by $\epsilon_f^{(\alpha)}=\sqrt{\alpha}(M+d/2)$.  
Therefore, for {\em any} flavour of exclusion statistics, the zero-temperature ODM  becomes

\begin{eqnarray} \label{heavy}
\rho^{(d)}(r, s) &=& \frac{g_s}{\pi^{d/2}}\sum_{n=0}^{\infty}\sum_{k=0}^{\infty} (-1)^n L_n^{d/2-1}(2r^2)L_k^{d/2-1}(s^2/2)e^{-(r^2+s^2/4)}\bigg[\frac{1}{\alpha} \Theta (\epsilon_f^{(\alpha)} - (\epsilon_n+k)) \bigg] \nonumber \\
&=& \frac{g_s}{\alpha \pi^{d/2}}\sum_{n=0}^{n_{\text{max}}}\sum_{k=0}^{k_{\text{max}}} (-1)^n L_n^{d/2-1}(2r^2)L_k^{d/2-1}(s^2/2)e^{-(r^2+s^2/4)} \nonumber \\
&=& \frac{g_s}{\alpha \pi^{d/2}}\sum_{n=0}^{n_{\text{max}}} (-1)^n L_n^{d/2-1}(2r^2)L_{k_\text{max}}^{d/2}(s^2/2)e^{-(r^2+s^2/4)}~,
\end{eqnarray}
where $n_{\text {max}} = \lfloor \sqrt{\alpha}(M+d/2) \rfloor -d/2$ and $k_{\text {max}}  = \lfloor \sqrt{\alpha} (M+d/2) \rfloor -n -d/2$.  Similarly,  the $T=0$ Wigner function reduces to
\begin{eqnarray} 
W^{(d)}(r, p) &=& 2^dg_s \sum_{n=0}^{\infty}\sum_{k=0}^{\infty}(-1)^{n+k}L_n^{d/2-1}(2r^2)L_k^{d/2-1}(2p^2)e^{-(r^2+p^2)} \bigg[\frac{1}{\alpha} \Theta (\epsilon_f^{(\alpha)} - (\epsilon_n+k)) \bigg] \nonumber  \\ 
&=& \frac{2^dg_s}{\alpha} \sum_{n=0}^{n_{\text {max}}}\sum_{k=0}^{k_{\rm {max}}}(-1)^{n+k}L_n^{d/2-1}(2r^2)L_k^{d/2-1}(2p^2)e^{-(r^2+p^2)}~ .
\end{eqnarray}
All of the zero temperature results in the fermionic sector may be obtained from Equations (41) and (42).

\section{Application}
While it may seem a little pedantic, we feel that it is useful to first illustrate the above discussion with a specific example, namely, the Gentile distribution, Equation (8).  Even though we {\em do not} require an explicit
expression for the thermal weighting factor, for the
GES, it can readily be shown that
\beq
\frac{H^{(\alpha_w)}(\beta;T)}{\beta} = - \frac{\pi T}{\tan (\pi \beta T)} +  \frac{1}{\omega}\frac{\omega\pi  T}{\tan \big(\omega\pi \beta T\big)}~.
\eeq
With the above form for the GES thermal factor, one may then work out  all of the two-sided ILTs explicitly, and readily confirm that this is equivalent to a direct
application of Eq.~(20), {\it viz.,}
\begin{eqnarray}
{\cal B}^{-1}_{\mu} \left[ e^{-(\epsilon_n+k)\beta} \frac{H^{(\alpha_w)}(\beta;T)}{\beta}\right] 
&=&
\frac{1}{\exp\big( \frac{\epsilon_n + k -\mu}{T} \big) -1} -  \frac{(\frac{1+\alpha_w}{\alpha_w})}{\exp\big((\frac{1+\alpha_w}{\alpha_w}) \frac{(\epsilon_n + k -\mu)}{T} \big) -1} \nonumber \\
&=&n^{(\alpha_w)}(\epsilon_n+k)~.
\end{eqnarray}
Note that at $T=0$, the fermionic branch ({\it i.e.,} $0<\alpha_w\leq 1$) of the GES becomes
\beq
n^{(\alpha_w)}(\epsilon_n+k) \to  \frac{1}{\alpha_w} \Theta (\epsilon_f^{(\alpha_w)} - (\epsilon_n+k))~,
\eeq
as mentioned above.
We therefore have closed form, analytical expressions for the one-body density matrix, and Wigner Function for GES, given by Eq.~(25) and (28), respectively, provided we take $n^{(\alpha)} \rightarrow n^{(\alpha_w)}$.

Let us continue the application of our results by also presenting the finite-temperature spatial density profiles (Eq.~(31)) for a harmonically trapped system obeying GES, and 
comparing them to those obtained from the FES distribution.   In this comparison, we identify $\alpha_w$ with the FES parameter $g$.~\cite{gentile,wang,nanda}
Our motivation is two-fold.  First we wish to illustrate the quality of the much simpler
GES distribution when evaluated for a local quantity (as opposed to integrated thermodynamic quantities such as the chemical potential, specific heat or energy per particle~\cite{local1,tanatar}).  
In addition we would also like to examine how differences in the spatial
densities may be used to probe the type of statistics exhibited by the system experimentally.  In particular, we have in mind applications to ultra-cold, harmonically trapped Fermi systems in the unitary regime
where there is suggestive evidence that the strongly interacting gas may be mapped to a noninteracting system obeying ideal FES.~\cite{bhaduri1,bhaduri2,bhaduri3,vanzyl_hutch,qin,anghel}

\begin{figure}[ht]
\begin{center}
\includegraphics[scale=.8]{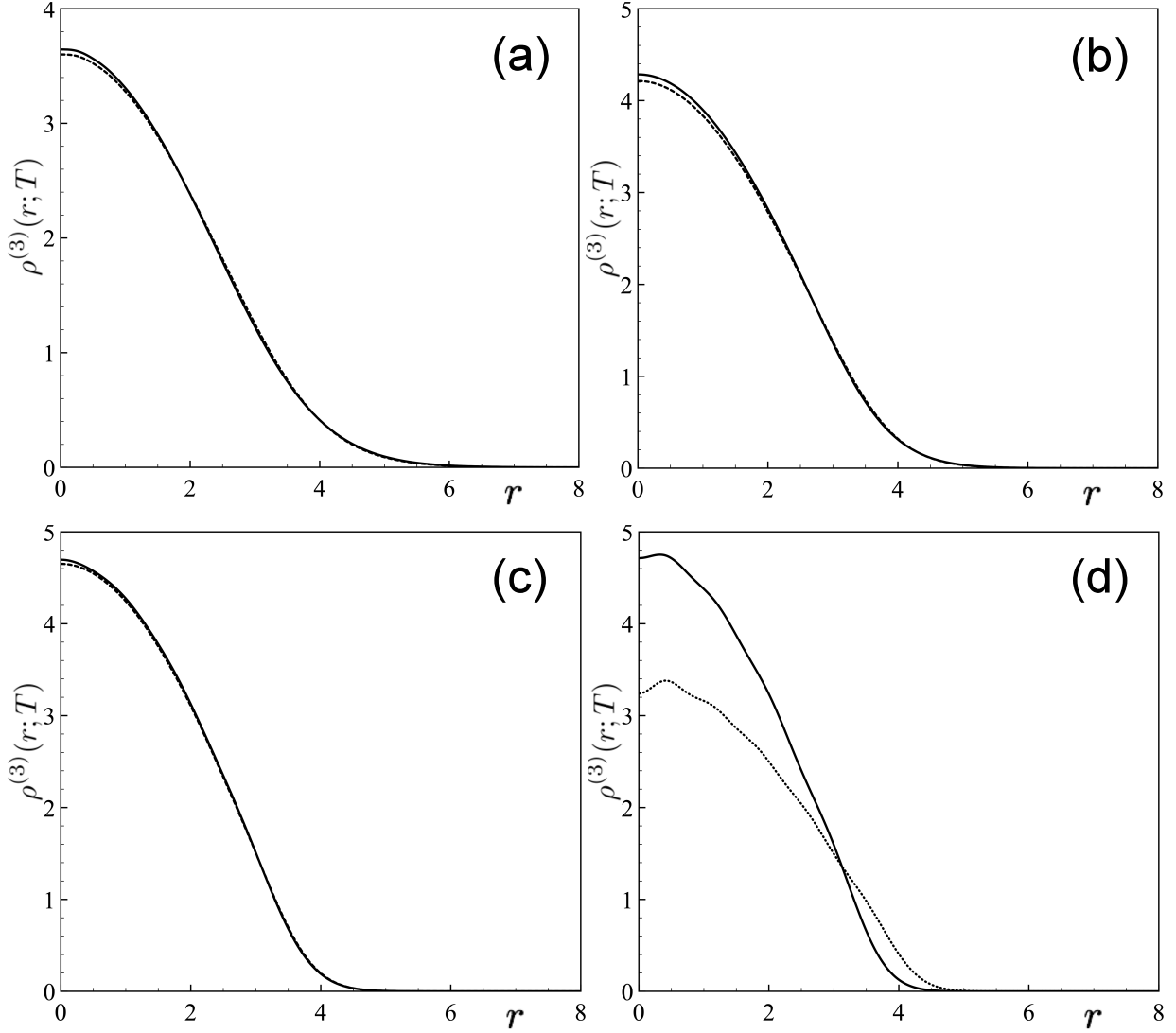}
\caption{Three-dimensional, finite temperature spatial density profiles for $N=420$ particles with $g=\alpha_w=1/2$ and $g_s=2$.  (a) $T=3$, (b) $T=2$, (c) $T=1$, (d) $T=0$.  
Solid curves are FES, dashed curves are GES, and the dotted line in panel (d) is the $T=0$ Fermi density.  Units are such that $k_B=\hbar=m=\omega=1$ as discussed
in the text.}
\end{center}
\end{figure}

In Fig.~1, we present the 3D spatial density profiles obtained from the GES (dashed curves) and FES (solid curves) at various temperatures, with $g=\alpha_w=1/2$ (semions).  Note that for this particular
value of the statistical parameter, an explicit form for the distribution function is available, and is given by Equation (5).  We note that at high temperature (panel (a)), the FES and GES spatial densities are
in good agreement, with the only significant deviation occurring near the center of the trap.  As the temperature is lowered, the agreement between the two densities improves.  In the zero temperature limit, the two spatial densities are analytically identical,
and the quantum mechanical shell oscillations become more prominent.  
As a reference, we have also included in the $T=0$ plot (panel (d)) the spatial density for the
Fermi statistics (dotted line, $\alpha_w=g=1$).  We observe that the smaller statistical parameter also effectively serves to ``bosonize'' the particles, resulting in an increase in the density at the origin, a squeezing of the distribution
in the tail region, and the diminished shell oscillations.

\begin{figure}[ht]
\begin{center}
\includegraphics[scale=.8]{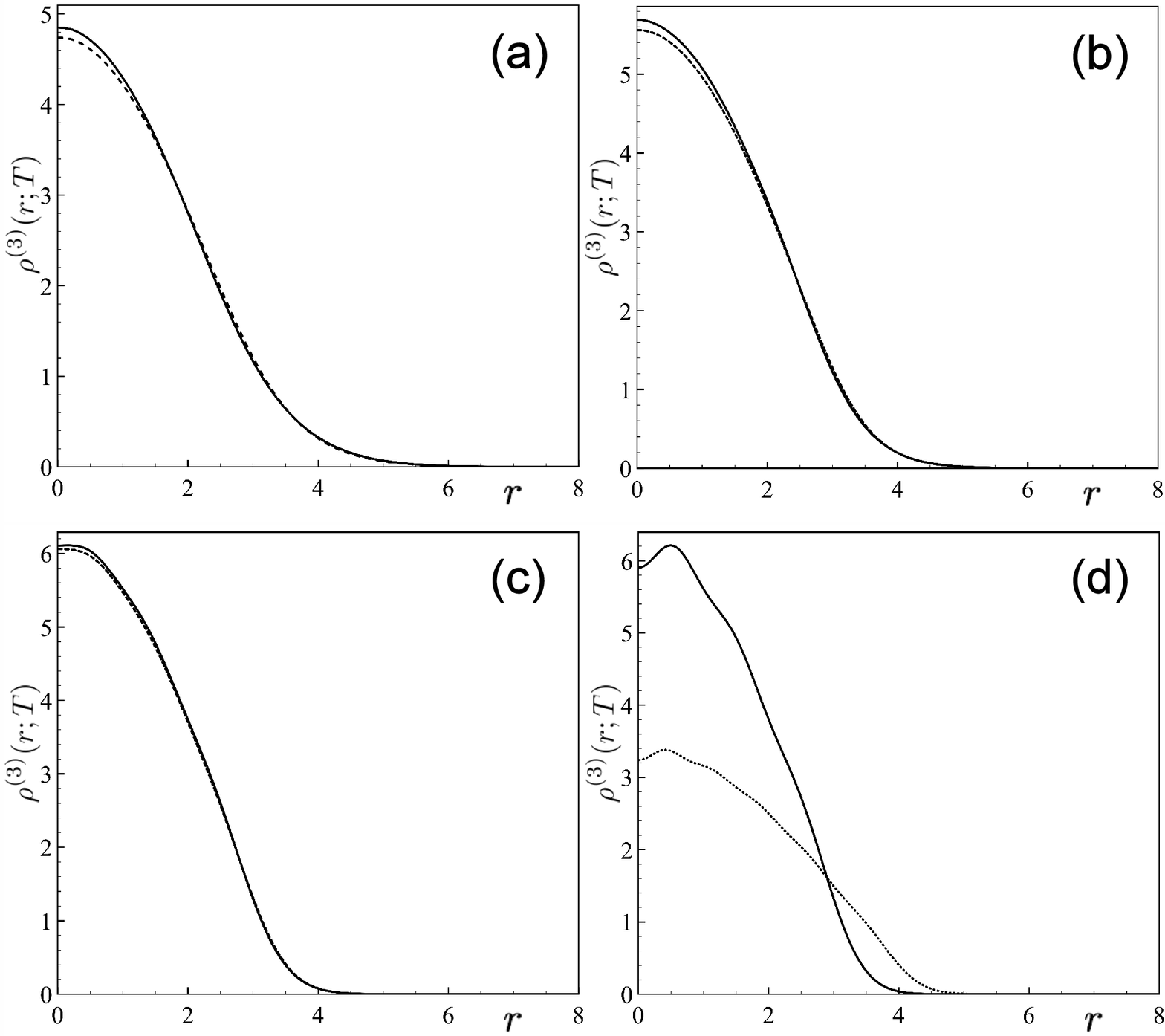}
\caption{ As in Fig.~1. $g=\alpha_w=0.29$.}
\end{center}
\end{figure}

In Fig.~2, we again present the spatial densities,  
but now with $g=\alpha_w=0.29$.  Our motivation for choosing this particular value of the statistical parameter lies in the earlier work of Bhaduri {\em et al}~\cite{bhaduri3} in the context of a harmonically
trapped ultra-cold Fermi gas, where it was argued that in the unitary regime, the strongly interacting system may be mapped onto a  gas of
particles obeying ideal FES.   In  their investigation, the statistical parameter was determined by fitting the finite temperature, theoretical FES energy per particle, $E/N$, and
chemical potential to the experimental data, thereby obtaining a ``best fit'' value of $g=0.29$.  
It is important to note, however, that by fitting to only global quantities, local information is
not included, which may be important in comparing to experimental data.  We suggest that by also examining the local spatial density, one may be able to provide further evidence
in support of the conjecture that the unitary Fermi gas obeys fractional statistics.

Figure 2 once again illustrates that  the GES and FES distributions are in good
agreement, and as the temperature is lowered, the agreement improves.  As in Fig.~1, the zero temperature limit results in identical spatial densities between the two distributions.  It is also clear that the reduced
statistical parameter leads to more boson-like behaviour of the particles, as evidenced by the significant squeezing of the cloud, and the increased density in the central region of the
trap, 
especially when contrasted with the Fermi density (panel (d), dotted line).    It would therefore be interesting to examine the experimental density distribution of the trapped Fermi gas in the unitary regime, and compare it to the theoretical predictions
of the GES and FES with $g=\alpha_w=0.29$ presented here.  
Owing to the similarities in the spatial distributions at low temperatures however (the regime of ultra-cold gases), it is unlikely that
one would be able to determine the specific kind of the statistics obeyed by the particles.   Nevertheless, the spatial density may well serve as another 
``smoking gun'' signature that the system in the unitary regime is indeed exhibiting fractional statistics.  

We would like to further mention that the agreement between the FES and GES extends to all values of the statistical parameter, and is also evident in other thermodynamic
quantities, which we have not detailed here.

\section{Closing Remarks and Conclusions}

We have presented closed form, universal analytical expressions for the finite temperature one-body density matrix, and Wigner function of a $d$-dimensional, 
harmonically trapped gas, obeying general exclusion
statistics.  These expressions, Eqs.~(25) and (28), completely generalize results presented elsewhere, which were limited to Bose and Fermi statistics.~\cite{vanzyl,vanzyl_wigner}
The universal forms of the one-body density matrix 
and Wigner function are only possible provided the normal Bloch-density matrix has its $\beta$-dependence
written in a purely exponential form. As a result, we have established that explicit knowledge of the thermal weighting factors, $H^{(\alpha)}(\beta;T)/\beta$, previously thought to be necessary,~\cite{vanzyl_bhaduri,vanzyl,vanzyl_wigner} for the evaluation of the one-body density matrix, are in fact {\em not} required.

As an application of our results, we have examined the GES distribution, Eq.~(8), which has recently been rediscovered,~\cite{wang} and  proposed as an alternative 
to the more complicated distribution found independently by Ramanathan and others.~\cite{ramanathan,ouvry,wu,isakov}
Through an examination of the local spatial density at finite temperature, we were able to demonstrate that the GES is a good description of the harmonically confined gas obeying FES.  Indeed, we have established that the low temperature ($k_BT/\hbar\omega<1$) global and local thermodynamic
properties derived from FES and GES are essentially indistinguishable.  As a result, we note that it would be unlikely to experimentally ascertain  the specific
underlying fractional statistics of an ultra-cold Fermi gas in the unitary regime, as all such distributions will tend to lead to the same low temperature properties.  In particular, any
suggestion that the unitary Fermi gas obeys ideal FES is somewhat arbitrary, as almost identical results will be found at
low temperatures using some other exclusion statistics distribution which smoothly interpolates between Bose and Fermi statistics.  
For example, while we have not presented the details here, we have confirmed that the spatial densities obtained
from the distribution given by Eq.~(15) are indistinguishable from GES and FES at low temperatures, although noticeable differences from GES and FES 
do occur at higher temperatures. 

Given the excellent agreement between the GES and FES distributions at low temperatures, we conclude that the much simpler GES may be used with confidence in other studies where simple, analytical
results for ultra-cold gases obeying FES are desired.
\acknowledgments
BVZ  would like to acknowledge financial support from the Discovery Grant program of the Natural Sciences and Engineering Research Council of Canada (NSERC).  ZM acknowledges additional
funding through the NSERC USRA program.  We would also like to thank Prof. M. V. N. Murthy for useful comments during the preparation of the manuscript, and for bringing
Refs.~\cite{ramanathan, nanda} to our attention.



\begin{thebibliography}{99}

\bibitem{haldane}
F. D. M. Haldane, Phys. Rev. Lett. {\bf 67}, 937 (1991).

\bibitem{murthy}
An excellent discussion of FES by M. V. N. Murthy and R. Shankar may be found at: http://www.imsc.res.in/~murthy/Papers/fesmu.ps

\bibitem{khare}
A. Khare, {\em Fractional Statistics and Quantum Theory}, 2nd ed. (World Scientific Publishing, Singapore, 2005).


\bibitem{ramanathan}
R. Ramanathan, Phys. Rev. D {\bf 45}, 4706 (1992).

\bibitem{ouvry}
A. Dasin\'eres de Veigy and S. Ouvry, Phys. Rev. Lett {\bf 72} , 600 (1994).

\bibitem{wu}
Y-S Wu, Phys. Rev. Lett. {\bf 73}, 922Ð925 (1994).

\bibitem{isakov}
S. B. Isakov, Phys. Rev. Lett. {\bf 73}, 2150Ð2153 (1994).

\bibitem{aring}
A. K. Aringazin and M. I. Mazhitov, Phys. Rev. E {\bf 66}, 026116 (2002)

\bibitem{gentile}
G. Gentile, Nuovo Cim. {\bf 17}, 493 (1940); Nuovo Cim. {\bf 19}, 109 (1942).

\bibitem{wang}
Q.  A. Wang, A. Le Mehaute, L. Nivanen, M. Pezeril, Nuovo Comento B {\bf 6}, 635 (2003).  

\bibitem{nanda}
V. S. Nanda, Proceedings of the National Institute of Sciences of India: Physical sciences {\bf  19}, 595 (1953).

\bibitem{sarkar}
K. Byczuk, J. Spalek, G. S. Joyce, and S. Sakar, Acta Physica Polonica B {\bf 26}, 2167 (1995)

\bibitem{note1}
The Gentile statistics also implements Eq.~(3) in an average sense, 
(with maximal occupancy of the state as $1/g$) without the correlations between states for occupancy rules. So Gentile statistics in this sense
also has elements of Haldane statistics, 50 years prior to Haldane's work in Ref.~\cite{haldane}.

\bibitem{brack_bhaduri}
M. Brack and R. K. Bhaduri, {\em Semiclassical Physics}, Frontiers in Physics, Vol.~96, Addison-Wesley, Reading, MA (2003).

\bibitem{polychrono}
A. P. Polychronakos, Phys. Lett. B {\bf 365}, 202 (1996).

\bibitem{vanzyl_bhaduri}
B. P. van Zyl, R. K. Bhaduri, A. Suzuki, and M. Brack, Phys. Rev. A {\bf 67} (2003).

\bibitem{grad}
I. S. Gradshteyn and I. M. Ryzhik, {\em Table of inegrals, series, and products}, $4$-th ed. Academic Press Inc., New York (1980).

\bibitem{shea_vanzyl}
P. Shea and B. P. van Zyl, J. Phys. A: Math. Theor. {\bf 40}, 10589 (2007).

\bibitem{vanzyl}
B. P. van Zyl, Phys. Rev. A {\bf 68}, 033601 (2003).

\bibitem{wigner}
 E. P. Wigner, Phys. Rev. {\bf 40}, 749 (1932).

\bibitem{vanzyl_wigner}
B. P. van Zyl, J. Phys. A: Math. Theor. {\bf 45}, 315302 (2012).

\bibitem{local1}
G. S. Joyce, S. Sarkar, J. Spalek , and K. Byczuk, Phys. Rev. B {\bf 53}, 990 (1996).

\bibitem{tanatar}
S. Sevincli and B. Tanatar, Phys. Lett. A {\bf 371}, 398 (2007).

\bibitem{bhaduri1}
R. K. Bhaduri, M. V. N. Murthy, and M. K. Srivastava, Phys. Rev. Lett. {\bf 76}, 165Ð168 (1996).

\bibitem{bhaduri2}
R. K. Bhaduri, M. V. N. Murthy, and M. Brack, J. Phys. B: At. Mol. Opt. Phys. 41 115301 (2008).

\bibitem{bhaduri3}
R. K. Bhaduri, M. V. N. Murthy, and M. K. Srivastava, J. Phys. B: At. Mol. Opt. Phys. {\bf 40} 1775 (2007).

\bibitem{vanzyl_hutch}
B. P. van Zyl and D. A. W. Hutchinson, 
Laser Physics Letters {\bf 5}, 162 (2008).

\bibitem{qin}
F. Qin and Ji-S Chen,  J. Phys. B: At. Mol. Opt. Phys. 43 055302 (2010).

 \bibitem{anghel}
 D-V. Anghel, preprint, arXiv:1204.0464v1



\end{thebibliography}
\end{document}